# Improving Electric Contacts to Two-Dimensional Semiconductors

Saurabh V. Suryavanshi, Blanka Magyari-Kope, Paul Lim, Connor McClellan, Kirby K.H. Smithe, Chris D. English, and Eric Pop

**Abstract**

Electrical contact resistance to two-dimensional (2D) semiconductors such as monolayer MoS$_2$ is a key bottleneck in scaling the 2D field effect transistors (FETs). The 2D semiconductor in contact with three-dimensional metal creates unique current crowding that leads to increased contact resistance. We developed a model to separate the contribution of the current crowding from the intrinsic contact resistivity. We show that current crowding can be alleviated by doping and contact patterning. Using Landauer-Büttiker formalism, we show that van der Waals (vdW) gap at the interface will ultimately limit the electrical contact resistance. We compare our models with experimental data for doped and undoped MoS$_2$ FETs. Even with heavy charge-transfer doping of $> 2 \times 10^{13}$ cm$^{-2}$, we show that the state-of-the-art contact resistance is 100 times larger than the ballistic limit. Our study highlights the need to develop efficient interface to achieve contact resistance of $< 10$ Ω·μm, which will be ideal for extremely scaled devices.

# I. Introduction

Two-dimensional (2D) semiconductors such as $MoS_2$ has shown a lot of promise for future transistor technology due to their sub-nanometer thickness and lack of out-of-plane dangling bonds. As the transistor channel lengths are scaled below 10 nm, we need a thinner channel to allow better gate control. In the case of 3D materials such as silicon, reduction in channel thickness below 4 nm causes drastic degradation in transport properties [1]. On the other hand, the layered nature of 2D materials with sub-1 nm monolayer (1L) thickness allows shorter channel length without degrading transport properties [2]. However, despite remarkable intrinsic properties such as mobility and reduced short channel effects, the performance of 2D field effect transistors (FETs) is strongly limited by the high contact resistance ($R_C$) [3]. Besides FETs, the electrical contact resistance is also a major concern for emerging applications of 2D materials such as spintronics [4] and valleytronics [5] that would require carrier injection from metal to 2D semiconductor. So far, contact resistance improvements to 2D materials have been achieved through ultra-clean contact metal deposition [2], doping [6], lithiation-induced phase change [7], , inserting insulators between metal and semiconductor (MIS) [8], and work function engineering [2],[9]. The best contact resistance, especially for 1L 2D materials, remains more than an order of magnitude higher (1 to 2 kΩ·μm or larger) than what is needed to demonstrate devices near their scaling limits (~50 Ω·μm) [3]. Such high contact resistances make it difficult to scale the FETs channel lengths as the device will be strongly dominated by the contact resistance [2].

To reduce the contact resistance further than the state-of-the-art, we need to understand the underlying physics. Recent works have focused on understanding the Fermi-level pinning in 2D materials [10] and understanding the carrier injection from metal to 2D materials [11]–[13]. Experimentally, researchers have tried to uncover the interface chemistry between metals and 2D materials [14]. Even though there is a growing interest, the literature so far fails to provide a holistic picture of electrical contacts to 2D materials.

In this paper, we provide a thorough understanding of the contact physics for 2D materials by isolating the extrinsic and intrinsic effects of the contact resistance. We achieve this by using a combination of *ab-initio* simulations, finite element simulations, fundamental limit calculations based on Landauer Buttiker formalism, and compact modeling. Such multiscale approach, as we will show later, is essential to get a complete picture of factors affecting contact resistance especially for novel technologies such as 2D materials. In addition to our multiscale simulation approach, we also fabricate transmission line method (TLM) structures to measure the contact resistance in real devices and compare these results with our simulations.

We quantify the current crowding based on the material thickness, carrier doping, contact geometry, carrier transport anisotropy, Schottky barrier height, and the "electrostatic doping" using back-gate. To further understand and optimize the current crowding, we propose two alternate contact architectures. We show that we can significantly reduce the contact resistance by patterning the contacts and managing the current crowding. In the second part, we calculate the fundamental (ballistic) limits of contact resistance for 2D materials. Using the transfer matrix method (TMM), we develop a model for carrier transmission at the contacts. We compare our simulation results with the experimental measurements and use the understanding to provide guidelines to further reduce the contact resistance below 50 Ω·μm

## II. Results and Discussions

The effective contact resistivity at the metal-semiconductor interface can be defined as a summation of the intrinsic contact resistivity ($\rho_i$) and the extrinsic effects such as the current crowding (C). The effective contact resistivity [15] is given as

$$\rho_C = \rho_i + Ct_{2D}\rho_{semi} \quad (1)$$

Here, $t_{2D}$ is the thickness of the channel (~ 0.5 nm to 6 nm) and $\rho_{semi}$ is the resistivity of the channel in units of Ω·μm. The current crowding (C) is defined as the additional resistivity faced by carriers before they are collected at the metal-semiconductor interface. This parameter 'C' captures the two-dimensional crowding below the contact, which is in addition to 1D crowding considered by the transmission line model (TLM). The effective contact resistivity allows us to use the TLM model to calculate the contact resistance $R_C \approx [\rho_C R_{sh}]^{1/2}$, where $R_{sh}$ is the sheet resistivity below the contact. Eq. (1) is a simplification to help us understand the relative contribution of the intrinsic and the extrinsic effects. We use a multiscale approach to uncover the contact physics and understand Eq. (1) in greater detail. See the methodology section for additional details.

Figure 1a shows a typical field effect transistor (FET). The back-gate is 5 nm SiO$_2$. We keep back-gate open-circuited (not connected to a voltage source) unless specified. The n-type background doping density is assumed to be $10^{19}$ cm$^{-3}$, which corresponds to $6\times10^{11}$

cm$^{-2}$ for a 1L MoS$_2$ channel. For a device such as a FET, the extrinsic resistance (or "contact") resistance) includes terms other than the metal-semiconductor contact resistance. Specifically, in a top-gated FET $R_{ext} = R_C + R_U$, where $R_C$ is the metal-semiconductor contact resistance and $R_U$ is the resistance due to the underlap (see inset of Fig. 1a). For this work, we will only be analyzing the metal-semiconductor contact resistances ($R_C$). We note that the metal being a 3D material does not necessarily form an intimate contact with the 2D material. As shown in the inset of Fig. 1a, the metals such as gold are more likely to form granular and non-intimate contacts. In addition, a typical metal-semiconductor interface will likely have impurities from the fabrication process. For now, we will neglect these non-idealists to understand current crowding. A FET is used as an example to study the electrical contact resistance. However, this work is relevant to any device that employs carrier injection between a metal and a 2D material.

### A.  Current Crowding in 2D Contacts

Current crowding has become a concern for modern devices as we increase the drive currents in the transistors and reduce the channel thickness (such as FinFETs and 2D devices). Current crowding has been extensively studied before for silicon transistors [15], [16]. There have been very few studies to understand current crowding in 2D devices [17], [18]. These studies have focused mostly on an empirical understanding of the current crowding in 2D materials and do not quantify the impact of channel thickness on current crowding.

In Fig. 1b, we plot the contact resistance calculated by assuming an ideal interface with the interface contact resistivity, $\rho_i = 0$. As per the TLM model ($R_C \sim [\rho_i R_{sh}]^{1/2}$), the expected contact resistance for such an ideal interface is zero. However, the carriers will face additional resistance from the semiconductor while entering the metal contact. We refer this phenomenon as current crowding. Because of this two-dimensional current crowding, we compute non-zero contact resistances as shown by blue symbols. We represent this two-dimensional current crowding using an effective contact resistivity ($\rho_C$) as shown in Eq. 1. This allows as to use the TLM equation $R_C = \rho_C/L_T \cdot \coth(L_C/L_T)$. $L_C$ is the total contact length, $L_T = [\rho_C/R_{sh}]^{1/2}$ is the current transfer length, and $R_{sh}$ ($\Omega/\square$) is the sheet resistivity of the 2D semiconductor below the contact. Fitting C to simulations, we observe $C \sim 0.18 \times [1 - \exp(-t_{2D}/L_T)]^{1/2}$. Note that since the carrier density and the mobility does not change below the contact, $R_{sh}$ remains unchanged due to crowding.

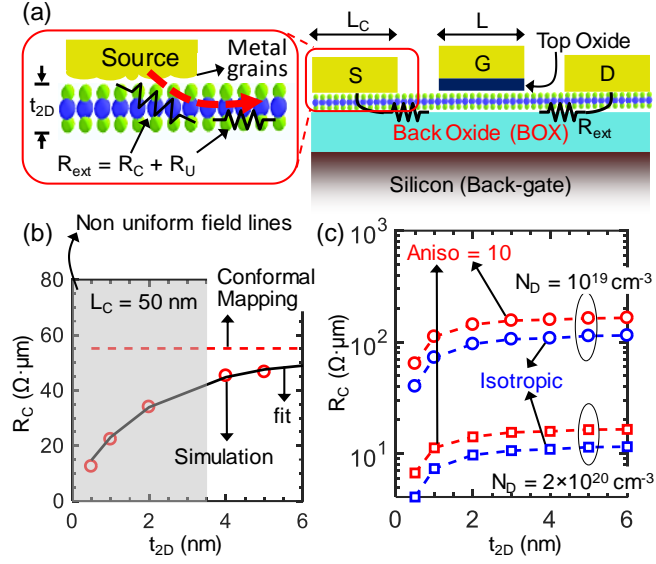

**Figure 1: (a) Contact resistance model:** The cartoon of a typical 2D FET is shown. The distributed model of the total external contact resistance for a FET ($R_{ext}$) is shown in the inset. **(b) Modeling current crowding**. We plot the simulated contact resistance (symbols) for different channel thickness ($t_{2D}$). The black lines are TLM model obtained by fitting the current crowding factor, C. The red dashed lines are the theoretical estimation of current crowding using conformal mapping. For TCAD simulations, we assume an n-type doping density $N_D = 10^{19}$ cm$^{-3}$ and carrier mobility of 50 cm$^2$V$^{-1}$s$^{-1}$. **(c) Impact of doping and transport anisotropy**. The contact resistance extracted from TCAD simulation is shown for two different background doping densities $N_D = 10^{19}$ cm$^{-3}$ (circle symbols) and $N_D = 2\times10^{20}$ cm$^{-3}$ (square symbols) and for isotropic transport (blue) and anisotropic transport (red, $\rho_{out}/\rho_{in} = \mu_{in}/\mu_{out} = 10$).

The red dashed lines in Fig. 1b plots the $R_{conformal}$ (derived in Eq. 7, methodology section) for different channel thicknesses. The $R_{conformal}$ is larger than the extracted from TCAD simulations because in conformal mapping we assume that the entire region below the contact is conducting which is not true in a real device. (Only the contact length of ~$3L_T$ participate in conduction.) Another discrepancy occurs as $t_{2D} \to 0$ because current flow lines are not uniform as assumed by the conformal theory. Nonetheless, as seen from Eq. (2), the contact resistance shows a dependence on geometry ($t_{2D}/L_T$) as expected from the TCAD simulations.

**Transport anisotropy and doping**. The 2D materials tend to exhibit transport anisotropy; that is, due to the layered nature of these materials, the in-plane mobility ($\mu_{in}$) is usually larger than the out-of-plane mobility ($\mu_{out}$). This anisotropy is emulated in the finite element simulations by a ratio of mobilities, which is equivalent to the ratio of bulk resistivities $\mu_{in}/\mu_{out} = \rho_{out}/\rho_{semi}$. Note that $\rho_{semi}$ and $\rho_{out}$ are in-plane and out-of-plane bulk resistivities of the 2D channel and have units of $\Omega \cdot \mu m$. Fig. 1c shows contact resistance for isotropic ($\mu_{in}/\mu_{out} = 1$) and anisotropic transport ($\mu_{in}/\mu_{out} = 10$). The in-plane mobility is kept constant during simulations $\mu_{in} = 50$ cm$^2$V$^{-1}$s$^{-1}$, which seems reasonable for state-of-the-art 2D materials such as MoS$_2$. As for the previous case, we assume an ideal interface with $\rho_i = 0$. With lower $\mu_{out}$ the resistance faced by the carriers entering the contacts is large, resulting in larger two-dimensional current crowding. As such, we observe that the contact resistance and therefore current crowding increases proportionally with the anisotropy or $C \propto \rho_{out}/\rho_{in}$. In Fig. 1c, we also increased the background n-type doping of the channel from $10^{19}$ cm$^{-3}$ to $2\times10^{20}$ cm$^{-3}$. The qualitative trend of C with anisotropy remains the same but we see a considerable reduction in the contact resistance due to reduced $\rho_{semi}$. This underlines the need to develop efficient and stable doping methods for 2D materials below the contact to reduce the contact resistance.

**Schottky contact depletion and back-gate dependence**. So far, we have assumed an ideal interface with an Ohmic contact. However, in real devices, the contact between an undoped semiconductor and a metal tends to be Schottky with a non-zero interface contact resistivity. In such a contact, due to the presence of the Schottky barrier at the interface, carriers below the contact may get depleted or accumulated depending on the applied voltage bias to the drain or the gate. In Fig. 2(a), we show the contact resistance for a Schottky barrier height from 20 meV to 100 meV. For Schottky contacts, the contact resistance has a dependence on the $V_{DD}$ and source and drain contact resistances are asymmetric. The calculated $R_C$, therefore, is an average value of both contact resistances. We apply a small drain voltage ($V_{DS} = 0.1$ V) to avoid introducing significant non-linearity in the contact resistance. Even with such small perturbations (small barrier height and small drain voltage), the TLM model used to extract the $R_C$ will show an error and is shown in Fig. 2a-b in the form of error bars.

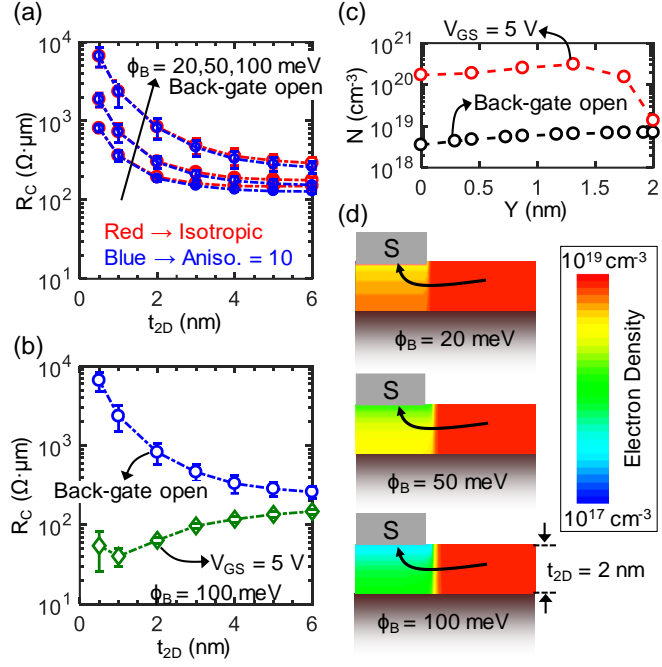

**Figure 2: Impact of Schottky depletion and back-gating on the current crowding.** (a) The extracted contact resistance for various Schottky barrier heights: $\phi_B$ = 20 meV, 50me, and 100 meV. The anisotropic transport is shown with blue symbols while isotropic transport is shown with red symbols. Note that TLM extractions for Schottky contacts have an error associated which are shown by error bars. (b) Th impact of back gate ($t_{BOX}$ = 5 nm of $SiO_2$, $V_{GS}$ = 5 V) in reducing the current crowding. (c) The electron density in the channel ($t_{2D}$ = 2 nm) below the contact for open-gate (black symbols) and $V_{GS}$ = 5V (red symbols). The background n-type doping density is $10^{19}$ cm$^{-3}$. (d) The electron (carrier) density below the contact without applying back-gate is shown for different barrier heights. The carrier density reduces drastically as the barrier height at the contact is increased. Here, $t_{2D}$ = 2 nm, $L_C$ = 50 nm. Since the semiconductor thickness is much smaller than the contact dimensions, we have re-scaled the figure disproportionately in x and y-direction to show the electron density in the semiconductor. We assume an n-type doping density $N_D$ = $10^{19}$ cm$^{-3}$, the in-plane mobility of 50 cm$^2$V$^{-1}$s$^{-1}$, and mobility anisotropy of 10.

As seen in Fig. 2a, $R_C$ is larger than what we observe in Fig. 1a mainly because of $\rho_i$ > 0 even for a small barrier height of 20 meV. As we reduce the channel thickness, the $R_C$ increases for all the Schottky barrier heights. Especially, it is worth noting that the effect of transport anisotropy is not significant (red and blue lines in Fig. 2a almost overalls). This is because crowding is dominated by the reduced carrier density below the contact and not by the anisotropic mobility. In Fig. 2b, we use the back-gate voltage ($V_{GS}$ = 5 V) to increase the carrier density below the contact. The application of back-gate voltage changes the carrier density distribution by attracting more carriers towards the semiconductor-BOX interface [18]. Though this increases current crowding (more carriers will have to travel larger distance before they reach the contact, shown in Fig. 2c), the overall $R_C$ is smaller due to a reduction in the contact depletion width. For even thicker semiconductor channels, the impact of increased current crowding might out-weigh the reduction in the contact depletion width and the contact resistance might be worse on applying back-gate voltage. However, such thicker channels ($t_{2D}$ > 6 nm) are not considered in this study. In Fig. 2d, we show the carrier (electron) density below the contact for different barrier heights. As we increase the barrier height, the carriers below the contact get depleted and therefore this results in higher contact resistance. Since this carrier depletion below the contact affects the crowding 'C' as well as $R_{sh}$, we could not quantify the 'C' parameter. But empirically we observe $C \propto \phi_B^{1/2}/t_{2D}$.

In experiments, $R_C$ tends to increase as the channel thickness is reduced [2], however our simulations do not capture this effect. Firstly, in simulations, we assume that the carrier mobility below the contact remains unaffected. On the other hand, in experiments, it is quite likely that metal deposition degrades the transport properties of the semiconductor below the contact [2]. This is especially true for a few layers of semiconductor and ends up increasing the contact resistance. Secondly, the 2D material properties such as band gap are thickness dependent. As a result, the Schottky barrier heights and other interface properties are different for different channel thickness. Thirdly, we point out that TCAD does not necessarily capture the interaction between metal and semiconductor. This is critical for contacts made to 1L or 2L materials. It has been shown that the metal wavefunction significantly modulates the *semiconducting* properties of the 2D material below the contact [19]. Lastly, the true nature of carrier injection from semiconductor would vary from device-to-device. In a device with an ideal contact, the only feasible ways for the carriers to inject from metal to semiconductor is at the contact edge [13]. However, in experimental devices with metal-semiconductor interaction and other non-idealities, we are more likely to have a top injection. Please see Appendix B for additional discussion.

**Reducing current crowding by contact patterning.** As seen from the previous discussion, we showed that it is possible to optimize the two-dimensional current crowding by doping, reducing the transport anisotropy, reducing the Schottky barrier height, and applying back-gate voltage. We also propose reducing the current crowding and contact resistance by using alternate contact architectures. In an extreme case, if we have an edge-only contact there would be no current crowding i.e. all the carries will enter the contact directly. Even though the edge contact is predicted to perform better than the top contact [20], the experimental results for the edge contact for 2D materials have so far been sub-par

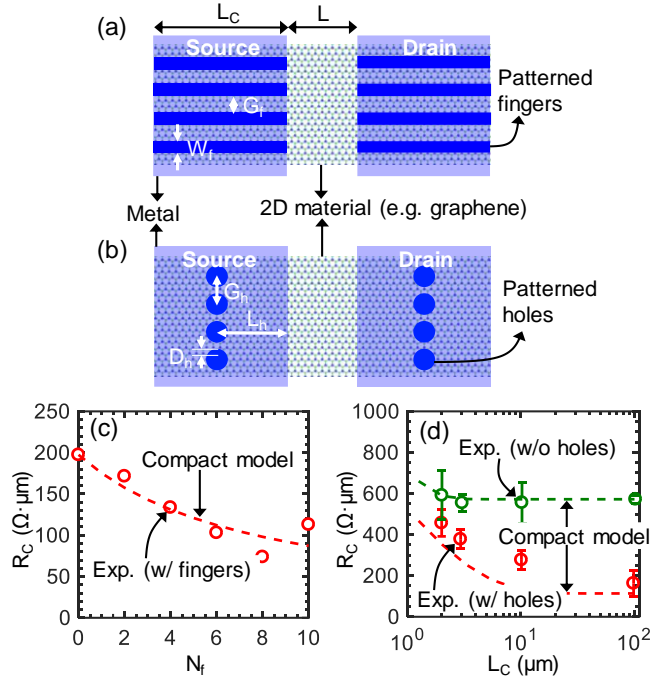

**Figure 3: Contact patterning to reduce the current crowding.** (a) Schematic of patterned finger contact architecture. (b) Schematic of patterned holey contact architecture. The 2D materials forms the channel. The contacts are shown in the light blue color. The patterned part which has semiconductor etched and then filled with metal is shown in dark blue color. We also show predictions of our compact model against the experimental results for 1L graphene (c) finger contacts and (d) holey contacts [22], [23].

[21]. It is possible to use a combination of "not-so-good" edge and top contacts to achieve a smaller contact resistance than either of edge-only or top-only contacts.

In Fig. 3, we showed two different contact architectures that can *manage* the current crowding; we pattern fingers (Fig. 3a) and holes (Fig. 3b) into the semiconductor below the contact. Both contacts have been experimentally shown for 1L graphene transistors [22], [23], though with no analysis and optimization. The size of fingers and holes in these experimental contacts are very large for applications to extremely scaled contacts. The pattern dimensions for optimized contacts, as we show, are closely related to the current transfer length or $L_T$, which depends on the sheet resistivity below the contact ($R_{sh}$) and the contact resistivity ($\rho_C$). Nevertheless, it is possible to achieve patterns of sub-10 nm dimensions by using technologies such as self-assembly [24]. Here we provide a systematic analysis of these contact architecture and develop compact models to estimate and optimize the contact resistance. We further perform finite element simulations and provide a guideline to design optimized contacts.

**Compact model for patterned contacts and experimental verification.** We can use models developed in Eq. 6 and Eq. 7 to explain the experimental data for contacts to graphene transistors (Fig. 3). Smith *et al*. [23] have fabricated fingered contacts to graphene transistor. We extract the contact resistance from the total device resistance as $R_C = (R_{total} - R_{ch})/2$. From their edge-only contact (contact with 14 fingers), we extract the $\rho_{edge} = R_C/t_{2D} = 2.7 \times 10^{-2}$ $\Omega \cdot \mu m^2$ assuming $t_{2D} = 0.315$ nm for graphene. From the top-only contact, we extract the $\rho_{top} = 40.9$ $\Omega \cdot \mu m^2$ assuming a TLM model. Using these values and our compact model from Eq. 6, we can predict the reduction in the contact resistance as seen in the experiments (Fig. 3c). In experiments, for further increase in the number of fingers the authors observe an increase in the contact resistance (not shown here), which is likely because there is increased current crowding between the fingers as the space between the fingers ($G_f$) reduces. Song *et al*. [22] have fabricated holey contacts and they extract $\rho_{edge} = 2.2 \times 10^{-1}$ $\Omega \cdot \mu m^2$. Assuming a TLM model, we extract $\rho_{top} = 545$ $\Omega \cdot \mu m^2$. Even though the experimental contacts have multiple arrays of holes, we consider the contribution only from the first layer of holes closer to the channel. Our predicted values for the contact resistance using Eq. 7 match the experimental values very well (Fig. 3d). Both experimental results discussed above show an improved contact resistance by patterning the contacts further supporting our theory.

We further use finite element simulations to optimize the contacts and provide guidelines for experimental fabrication. Simulation helps us to capture the current crowding more accurately. To simplify optimization, we assume ohmic contacts. For all simulations we assume $t_{2D} = 5$ nm, $\rho_{top} = 100$ $\Omega \cdot \mu m^2$, $\mu = 50$ cm$^2$V$^{-1}$s$^{-1}$, and $N_D = 10^{20}$ cm$^{-3}$. This implies sheet resistivity $R_{sh} = 2500$ $\Omega/\square$ and the current transfer length $L_T = 200$ nm. We also keep the contact length $L_C = 200$ nm. The contact resistance due to the top-only contact (no patterning) is $R_{Ctop} = 650$ $\Omega \cdot \mu m$. The purpose of the patterning, therefore, is to achieve contact resistance smaller than 650 $\Omega \cdot \mu m$.

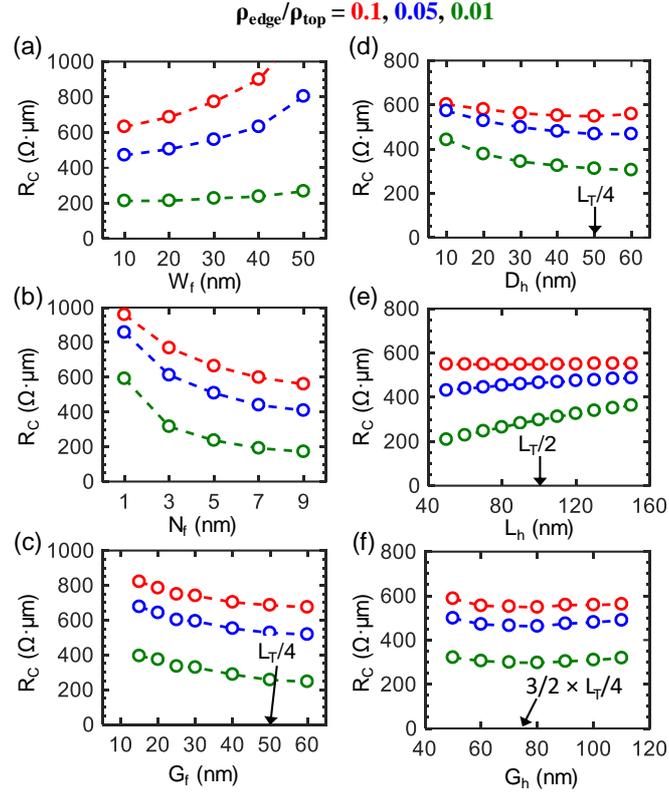

**Figure 4: Optimizing the fingered and holey contact using TCAD simulations.** Finger contact optimization for (a) finger width or $W_f$, (b) number of fingers or $N_f$, and (c) distance between fingers or $G_f$. Holey contact optimization for (d) hole diameter or $D_h$, (e) distance of holes from the edge or $L_h$, and (f) distance between holes or $G_h$. The device width for all simulation is 400 nm and the contact resistance are normalized per width.

**Optimizing fingered contact.** In Fig. 4a, we show the impact of the $W_f$ on the contact resistance keeping $N_f$ constant (= 5) and in Fig. 4b we show the impact of the $N_f$ keeping $W_f$ constant (= 10nm). If $\rho_{edge} > 0.1\rho_{top}$, the contact resistance is larger than the top-only contact resistance ($R_{Ctop}$) which is not desirable. From Fig. 4a, we observe that the smaller finger width provides smaller resistance. Thus, the optimized finger width will be limited by the lithographic limitations. From Fig. 4b, we observe that as we increase $N_f$, we reduce the resistance and it saturates after $N_f = 9$. This is because, as we further increase $N_f$ the distance between the fingers ($G_f$) reduces and increased current crowding mitigates the impact of the edge conduction. To further optimize for $G_f$, we keep $N_f = 5$ and $W_f = 10$ nm in Fig. 4c. We observe that the optimum $G_f$ is $L_T/4 = 50$ nm. With these optimized value, we could get the lowest contact resistance $R_C = 150$ $\Omega \cdot \mu m$ for $\rho_{edge} = 10^{-8}$ $\Omega cm^{-2}$, $\rho_{top} = 10^{-6}$ $\Omega cm^{-2}$, which corresponds to ~77% reduction in the contact resistance compared to the top-only contact.

**Optimizing holey contact**. In Fig. 4d-f, we optimize the holey contact. As seen for fingered contacts, the improvement in contact resistance is observed only if $\rho_{edge} < 0.1\rho_{top}$. We observe that as we increase the hole diameter ($D_h$) keeping the number of holes same, the resistance reduces and then saturates (Fig. 4d). The optimal $D_h$ is $L_T/4$, which is equal to 50 nm in this case. In Fig. 4e, we keep the $D_h$ = 50 nm and change the distance of the holes from the contact edge ($L_h$). If holes are too far away from the edge, the edge conduction does not help. If they are too close to the contact, we lose on the conduction from the top contact. We observe that the optimal $L_h$ is $L_T/2$ (=100 nm) from the contact edge. As seen in Fig. 4f, the optimal distance ($G_h$) between the holes is $3/2 \times L_T/4$ (= 75 nm). With these optimized values, we observe that the lowest $R_C$ ~ 250 $\Omega\cdot\mu m$ for $\rho_{edge} = 10^{-8}$ $\Omega\cdot cm^{-2}$, $\rho_{top} = 10^{-6}$ $\Omega\cdot cm^{-2}$, which corresponds to ~62 % reduction in the contact resistance compared to the top-only contact.

From our simulations, we observe that current crowding alone cannot account for a high contact resistance (> 1 k$\Omega\cdot\mu m$) observed in the literature. Further, it is possible to reduce the current crowding to extremely small value by contact patterning and therefore is not a fundamental limitation. Going back to Eq. 1, it is likely that that the contact resistance is limited due to the intrinsic contact resistivity. We explore this intrinsic contact resistivity in the next section.

## B. Fundamental (Ballistic) Limit for Contact Resistance to 2D Materials and Comparison with Experiments

Landauer-Büttiker formalism calculations for contact resistance have been previously done for 3D materials [25], [26] as well as most recently for 2D materials [27]. These studies assumed an ideal transmission which is difficult to realize physically. Here, we model the fundamental limits to contact resistances to 2D materials. We also fabricate TLM structures for 1L $MoS_2$ and compare our model with the measured results.

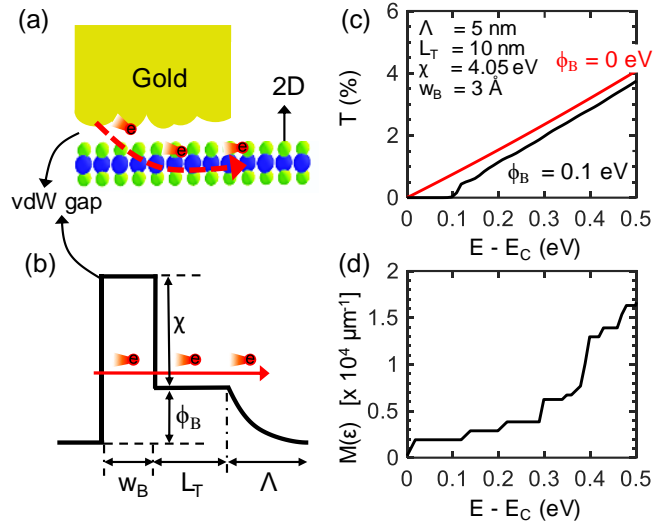

**Figure 5: Transmission and conduction modes at the metal-MoS$_2$ interface**. (a) Schematic to elaborate injection of a typical carrier at the metal-2D interface. (b) A typically potential profile observed by the carrier while traveling from metal to the channel. $\phi_B$ is the Schottky barrier height, $\chi$ is the electron affinity of the semiconductor below the channel, $\Lambda$ is the Schottky barrier penetration into the channel, and $L_T$ is the current transfer length or in other words, the average distance traveled by the carriers before they enter the channel. (c) Carrier transmission as a function of the energy using the transfer matrix method. The transmission is shown for two different barrier heights of 0 eV and 0.1 eV. The effective mass of the carriers is assumed to 0.45$m_0$, where $m_0$ is the free electron mass. (d) The number of modes for a 1L MoS$_2$ as a function of energy calculated using the dispersion relation for 1L MoS$_2$.

**Calculating modes and transmission**. In a real 2D FET, the carrier must pass through multiple barriers before it could reach the channel as shown in Fig. 5a and Fig. 5b. Because the metal deposition is not uniform, and metal is likely to not have an atomically smooth surface the van der Waals (vdW) gap between the contact metal and the channel semiconductor is going to change as shown in Fig. 5a. To simplify the calculation, we will assume an effective vdW gap with width $w_B$. The carrier travels an average distance proportional to the transfer length ($L_T$). The transmission coefficient (Tr) is then calculated using the transfer matrix method. Figure 5c shows transmission for $\phi_B = 0$ eV, and 0.1 eV. Note that for energies below the barrier height the transmission is negligible i.e. all the carried are reflected into the contact. These carriers might be able to tunnel through the barriers if there are traps or metallic in the semiconductor. Since our model does not include the trap-assisted tunneling, we are slightly underestimating the transmission. However, the *fitted* barrier height can be treated as an effective barrier height at the metal-2D contact which includes the effect of traps. In Fig. 5d, we show $M_{2D}(\epsilon)$ calculated for single layer MoS$_2$ (Eq.

3). For about 100 meV into the conduction mode, the total number of modes available for conduction is ~2000/μm. This is comparable to the number of modes available in Silicon for a channel of similar thickness [25]. From Fig. 5d, we also note that the $M_{2D}(\epsilon)$ increases are we go higher in the conduction band as seen for 3D materials [25]. Using the $Tr(\epsilon)$ and $M_{2D}(\epsilon)$, we can calculate the contact resistance from Eq. 2.

**Comparison with experiments**. An atomic force microscopy image of a TLM structure on $MoS_2$ doped with anisotropic $AlO_x$ is shown in Fig. 6a. Using the TLM structure, we extract $R_C$ for different back-gate voltages (and carrier density) [additional information regarding contact resistance extraction can be found in Ref. [28]]. Fig. 6b shows calculated intrinsic $R_C$ assuming $Tr(\epsilon) = 1$ for single layer $MoS_2$ (black bold lines) along with $R_C$ extracted from experimental devices (red and black symbols). We notice that the fundamental (intrinsic) contact resistance for 1L $MoS_2$ is smaller than 10 Ω·μm at the carrier density of $10^{13}$ cm$^{-2}$ and reduces further for higher carrier densities.

In our experiments, $R_C$ decreases with the carrier density as seen by the blue curve. Doping the channel with sub-stoichiometric $AlO_x$ helps us to achieve even higher carrier density as seen by the red symbols in Fig. 6b [29]. We could experimentally achieve the lowest contact resistance of ~480 Ω·μm at a carrier density of $2\times10^{13}$ cm$^{-2}$. This contact resistance is still ~300x away from the fundamental limit. This carrier density is limited by the maximum electric field that can be achieved using our gate oxides, which is thermally grown $SiO_2$. It is worth noting that the charge density induced by doping (doped device) as well as a back-gate electric field (undoped device) have a similar impact on the $R_C$. By doing an extrapolation on the experimental $R_C$, we see that we can potentially achieve a contact resistance of 100 Ω·μm at a very high carrier density of $10^{14}$ cm$^{-2}$. Though such carrier density has been achieved by using ionic gel gates [30], so far, no CMOS-compatible doping technique has been able to achieve such high carrier densities. Even if it is possible to have such a high doping, we must be able to selectively dope the contacts without "spilling" over the doping to the channel and changing the channel threshold voltage. We, therefore, need to explore other avenues to achieve low $R_C$ especially at carrier densities of $10^{13}$ cm$^{-2}$ or below.

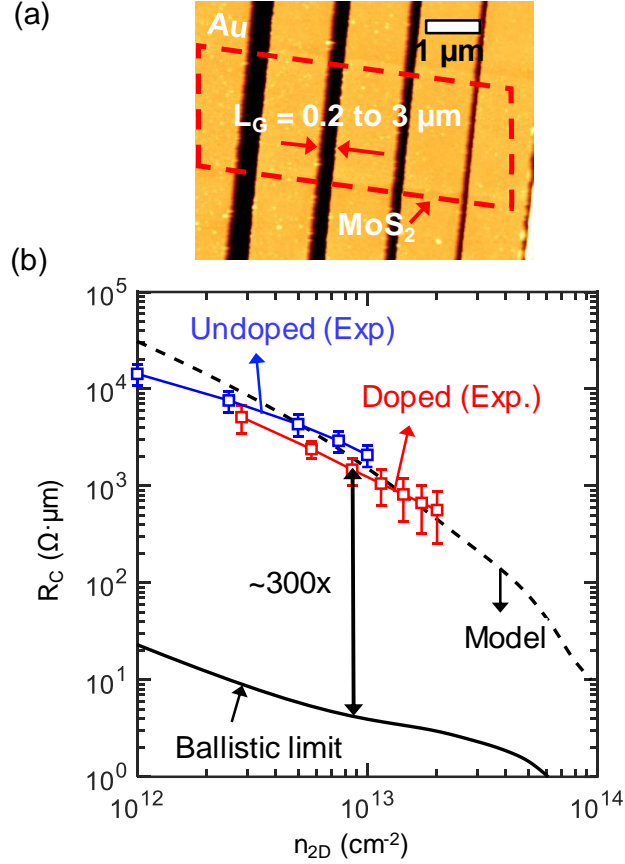

**Figure 6: Experimental measurements and comparison with intrinsic limits.** (a) An AFM image for TLM structure used to measure the contact resistance. (b) We compare the fundamental intrinsic contact resistance for MoS$_2$ with our experimental results for doped (red symbols) and undoped (blue symbols) MoS$_2$ contacts. The fundamental limit to contact resistance to MoS$_2$ is shown by the bold block line while our model fit is shown by the black dashed line.

We use the experimental data to fit our transmission matrix model. The black bold curve represents an ideal scenario with $Tr(\epsilon) = 1$ while the black dotted curve represents realistic transmission calculated using TMM and shown in Fig. 5c. For our device, we achieved a fit using $w_B = 3$ Å, $\phi_B = 50$ meV, $L_T = 10$ nm, $\Lambda = 5$ nm, $\chi = 4.05$ eV. Note that, among these parameters, the transmission (and the contact resistance) is most sensitive to $w_B$ and $\phi_B$. On the other hand, the transmission is not very sensitive to changes in $L_T$ and $\Lambda$. Our calculations suggest that to reduce the contact resistance further we need to reduce the $w_B$ and $\phi_B$. In other words, in order to reach the fundamental limit of <10 Ω·μm, we need an ohmic contact which is intimate and does not a have vdW gap.

# III. Conclusion

In summary, we have studied the electronic contact resistance to 2D materials using a multiscale approach including *ab-initio* simulations, finite element simulations, Landauer-Büttiker formalism, and compact modeling. We have confirmed our simulations results by fabricating TLM structures. We have shown that the metal-2D interface and not the current crowding is limiting the state-of-the-art contact resistance.

We have analyzed the contact resistance by separating the intrinsic components ($\rho_i$) and the extrinsic components (C). From our simulations, we have observed that the current crowding is not significant for extremely thin devices with $t_{2D} < 1$ nm. We have uncovered the current crowding due to contact geometry (C $\propto L_C/t_{2D}$), transport anisotropy (C $\propto \rho_{out}/\rho_{semi}$), and the Schottky barrier (C $\propto \sqrt{\phi_B}/t_{2D}$). We note that, especially for 1L channels, Schottky depletion will have the largest impact on the contact resistance. However especially for thin channels, $t_{2D} < 2$ nm, we have shown that the back-gate voltage improves the carrier density below the contact and reduces the current crowding.

To improve the contact resistance and reduce the current crowding, we proposed alternative (patterned) contact architectures to manage the current crowding. The different contact architecture can provide an improvement in contact resistance by reducing the current crowding provided that the edge contact is sufficiently good ($\rho_{edge} < 0.1\rho_{top}$). For fingered contacts, the optimized design should have extremely thin fingers (smallest $W_f$ limited by lithography) and should be separated from each other by ($G_f$) $L_T/4$. The optimized holey contacts should have holes of diameter ($D_h$) $L_T/4$. These holes should be separated from each other ($G_h$) by $3/4 \times L_T/2$ and should away from the contact edge ($L_h$) by $L_T/2$ for optimized contacts. These optimized contacts can provide a 77 % reduction in the contact resistance.

We performed full band calculations using Landauer-Büttiker formalism for fundamental limits of contact resistance to 2D materials. We have shown that it is possible to achieve a contact resistance of < 10 Ω·μm (at the carrier density of $10^{13}$ cm$^{-2}$) for 2D materials. The state-of-the-art contact resistance is 100 times higher than the fundamental limit. This larger contact resistance is primarily because the transmission at the metal-2D interface is extremely poor (i.e. the contacts have larger $w_B$ and $\phi_B$). As such, to achieve further improvement in contact resistance we have shown that it is necessary to focus on improving the metal-2D interface.

# IV. Methodology

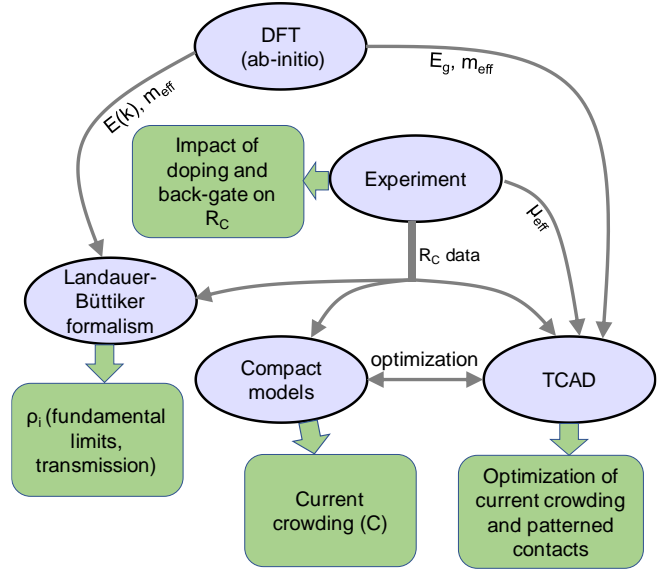

**Figure 7: Schematic of inter-dependence of various methodologies.** The blue ovals depict the main methodologies used in this study. The grey line shows the inter-dependence between different technologies and the flow of information. The green squares show the motivation for these individual methodologies. E(k) is the dispersion relation in the first Brilloin zone, $m_{eff}$ is the effective mass of electron, $E_g$ is the electronic bandgap, $\mu_{eff}$ is the effective mobility of the 2D material, and $R_C$ is the measured contact resistance for comparison

Seemingly different simulation methodologies are necessary to get a holistic picture especially in the case of novel materials such as 2D materials. For example, *ab-initio* simulations are necessary to calculate the material properties such as band gap and effective mass required for TCAD simulations. *ab-initio* simulations also provide a dispersion relationship required by Landauer-Büttiker formalism. The compact model depends on the TCAD simulations fit key parameters and optimize current crowding. In the end, we need controlled experimental results ($R_C$) to compare our simulations with. In addition to understanding the contact resistance for 2D materials, we also hope that this study will serve as a guideline for understanding contact resistances for future technologies. Figure 7 shows a schematic depiction of the inter-dependence of different methodologies used in this study.

## A. Theory

**Density Functional Theory (DFT)** (**or *ab-initio* simulation**): DFT is a computational quantum mechanical modeling method used to study the solid-state properties of materials such as electronic structure and dispersion relations. The properties of many-body electron systems are determined using a functional describing the electron density instead of positions of every electron. The Born-Oppenheimer approximation allows decoupling of adiabatic Schrödinger equations of electrons and nuclei. The nuclei produce a static electric field in which electrons move. The many-body Schrödinger is then approximated by self-consistently solving Kohn-Sham equations [31].

**Technology Computer Aided Design (TCAD) simulations**: TCAD simulations refer to finite element simulation methodology that self-consistently solve the Poisson and current continuity under the drift-diffusion approximation. The simulation region (the electrical device in this case) is divided into smaller regions by meshing and the calculations are carried using a finite element method.

**Landauer-Büttiker formalism**: In the limit of no current crowding (C = 0), the contact resistivity ($\rho_C$) is limited by the intrinsic contact resistivity ($\rho_i$). The $\rho_i$ for 2D materials is quantified using the Landauer-Büttiker formalism [25], [32]. As per this formalism, the intrinsic contact resistivity to 2D material in the ballistic limit is given as follows,

$$\frac{1}{\rho_i} = \frac{4q^2}{h} \int_{-\infty}^{\infty} M_{2D}(\varepsilon) Tr(\varepsilon) \left[-\frac{\partial f}{\partial \varepsilon}\right] d\varepsilon \qquad (2)$$

Here, q is the unit electronic charge, h is Planck's constant, f is the Fermi-Dirac distribution. $M_{2D}(\epsilon)$ is the number of transport modes as a function of energy and $Tr(\epsilon)$ is the transmission coefficient at the interface between the semiconductor and the metal. For a 2D channel, the units of $\rho_i$ are $\Omega \cdot \mu m$, which is equivalent to $R_C$. The metal is assumed to be ideal and having sufficient conducting modes so that the contact is limited by the modes in the semiconductor. The energy derivative of the Fermi-Dirac distribution $[-\delta f/\delta \epsilon]$ activates the modes within an energy of $\sim k_B T$ around the Fermi energy, where $k_B$ is the Boltzmann constant and T is the average interface temperature. Only these "activated modes" participate in conduction at the interface and contribute to the contact resistance. Eq. (2) can also be understood as the product of the quantum conductance ($2q^2/h$) and the number of available conducting modes per unit width of the semiconductor channel. An additional factor of 2 is used to account for two contacts.

The $M_{2D}(\epsilon)$ is calculated by counting the number of modes from the band structure as shown by the following formula [25],

$$M_{2D}(\varepsilon) = \frac{1}{2\pi} \int_{BZ} \sum_{n+} \Theta(\varepsilon - \varepsilon_{k_y}) dk_y \qquad (3)$$

Here, the modes are counted along the transport direction in the first Brillion zone. The unit of $M_{2D}(\epsilon)$ is $\mu m^{-1}$. $\Theta(\epsilon)$ is the dispersion relationship calculated using the DFT simulations. $k$ is the momentum vector. For a 2D system, the $M_{2D}(\epsilon)$ is proportional to $\sqrt{m_\perp}$ [25], where $m_\perp$ is the effective mass along the transverse direction or the direction perpendicular to the transport. So effectively, $R_C \propto 1/\sqrt{m_\perp}$. Since 2D materials are known to have higher effective masses, they are, therefore, expected to have a lower limit to the $R_C$.

**Conformal mapping**: The TCAD simulation is supplemented with a compact model for contact resistance exclusively due to current crowding. An analytical expression is derived in Eq. 5 for the contact resistance ($R_{conformal}$) in the limits of $\rho_i = 0$ using conformal mapping [33]. Such calculations were previously used to calculate resistance in thin film patterns [33] as well as fringe capacitances between non-parallel surfaces [34]. For these calculations, the device is first defined in a complex plane $z = x + iy$ as shown in Fig. 8a. Using an appropriate transformation function [33], this pattern is transformed into a more familiar pattern of a semiconductor bar and two metal contacts in the $w = u + iv$ plane as shown in Fig. 8b. The primary assumption of this transformation is that the flow lines are uniform and elliptical. The transformation function (f) can, therefore, be defined as $w = f(z)$. This function transforms the flow lines (current potential lines) from the z-plane to the w-plane. For the ease of visualization, one of the many flow lines is shown by the dashed red line in Fig. 8. In the z-plane the resistance is $R_z = 2R_C + R_{ch}$. Here, $R_{ch}$ in the channel resistance. For the limit, $L_W \to \infty$, $L \to \infty$, the contact resistance is calculated as,

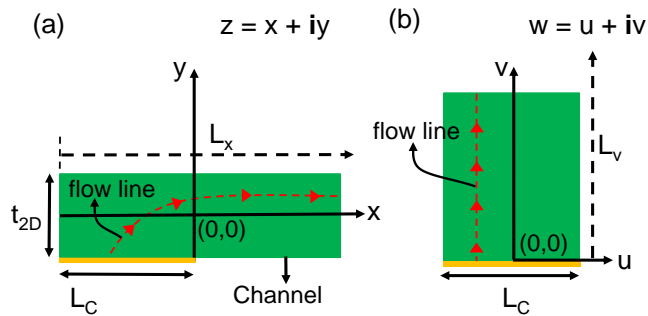

**Figure 8: Conformal Mapping**. (a) A depiction of a contact in a complex plane (z-plane). (b) After transformation, we convert the geometry into a simpler form where it is easy to calculate the resistance. The yellow line is the contact metal and the green area is the semiconducting channel. In both figures, the flow lines are shown by red dashed lines.

$$R_{Conformal} = \frac{1}{2} \lim_{L_w \to \infty, L \to \infty} [R_w - R_{ch}] \qquad (4)$$

This is further simplified to obtain an analytical expression for the $R_{Conformal}$ or the resistance due to current crowding. $L_C$ is the contact length and $t_{2D}$ is the thickness of the semiconducting channel.

$$R_{Conformal} = \frac{\rho_{semi}}{W} \left[ \frac{L_C}{t_{2D}} - \frac{2}{\pi} \log \left( \sinh \left( \frac{\pi L_C}{2 t_{2D}} \right) \right) \right] \qquad (5)$$

**Compact model for patterned contacts**: Two different compact models are developed here: contacts with etched metal fingers (fingered contacts) and contacts with etched holes (holey contacts). For fingered contacts (Fig. 3a), the fingers are assumed to have a uniform width ($W_f$) and are equally spaced (by $G_f$) from each other. The carriers can enter the metal contact through the side edge of the finger that is parallel to the current flow or the front edge of the finger that is perpendicular to the current flow. These flow paths are in addition to the carriers going directly to the top contact. The total contact resistance of such an architecture can be divided into three parallel contributions: top contact ($R_{top}$), the front edge contact ($R_{fe}$), and the side edge contact ($R_{se}$) such that $R_C = R_{top} \parallel R_{fe} \parallel R_{se}$.

$$R_{top} = \frac{1}{(W - N_f W_f)} \frac{\rho_{top}}{L_T} \coth \left[ \frac{L_C}{L_T} \right] \qquad (6a)$$

$$R_{fe} = \frac{\rho_{edge}}{N_f W_f t_{2D}} \qquad (6b)$$

$$R_{se} = \frac{1}{2 N_f t_{2D}} \frac{\rho_{edge}}{L_T} \coth \left[ \frac{L_C}{L_T} \right] \qquad (6c)$$

Here, $\rho_{top}$ is the top contact resistivity ($\Omega \cdot \mu m^2$), $L_C$ is the contact length, $L_T$ is the current transfer length, $t_{2D}$ is the semiconductor channel thickness, W is the channel width, $\rho_{edge}$ is the edge contact resistivity ($\Omega \cdot \mu m^2$), and $N_f$ is the total number of fingers patterned below the metal contacts.

For the holey contact (Fig. 3b), the current can either enter through the edge contact of the holes or through the top contact. The holes with a diameter of $D_h$ and are equally spaced by $G_h$. An additional parameter in holey contact architecture is the distance of the holes from the contact edge ($L_h$). Note that only half of the hole circumference facing the current flow will participate in the conduction. If there are additional arrays of the holes, they will be shadowed by the first row of holes closer to the channel and therefore will not be able to participate in the conduction. The total contact resistance for the holey contact can be divided as a parallel combination of the contact due to holes $R_h$ and top contact $R_{top}$ such that $R_C = R_h \| R_{top}$.

$$R_{top} = \frac{1}{W - N_h D_h} \frac{\rho_{top}}{L_T} \coth\left[\frac{L_C}{L_T}\right] \quad (7a)$$

$$R_h = \frac{1}{N_h}\left[\frac{R_{sh}}{2\pi} \cosh^{-1}\left(\frac{2L_h}{D_h}\right) + \frac{\rho_{edge}}{\pi D_h t_{2D}}\right] \quad (7b)$$

Here, $N_h$ is the total number of holes etched below the contact and $R_{sh}$ is the sheet resistivity below the contact.

### B. Computational Details

**DFT:** The total number of modes, $M_{2D}(\epsilon)$, as discussed in Section A is calculated from the electronic dispersion relation of the 2D materials in the first Brillion zone using *ab initio* simulations. These density functional theory simulations are performed using the Vienna *Ab-initio* Simulation Package (VASP) [35]. We use the projected-augmented wave method (PAW) [36] with Perdew-Burke-Ernzerhof (PBE) potentials for structure relaxation and employ hybrid functional HSE06 for accurate calculations of the dispersion relation and band gaps [37]. The van der Waals interactions are considered by the Tkatchenko-Scheffler dispersion correction method which has been previously used for 2D materials [38]. For our simulations, we use an energy tolerance of $10^{-6}$ eV per atom and a force tolerance of $10^{-2}$ eV/Å per atom for atomic relaxations. The simulations are performed with periodic boundary conditions on all three dimensions. In the direction perpendicular to the material surface, a vacuum of 1 nm is used to avoid interaction between neighboring cells. The K-point sampling for simulations is 40×40×1.

**TCAD**: The three-dimensional (3D) finite element simulations are performed using Sentaurus Device [39]. For contacts, whenever required, a Schottky barrier is assumed at the metal-2D semiconductor interface. A non-local tunneling model is employed to account for tunneling at the Schottky barriers, which considers the band profile along the entire tunneling path. This is especially required in the case of current crowding calculations where the electric field changes along the tunneling path [40]. The tunneling probability along this path is calculated using Wentzel-Kramer-Brillouin (WKB) approximation. Because of the layered nature of 2D materials, the in-plane mobility is larger than the out-of-the-plane mobility. The material properties of the 2D materials such as band gap (~ 2 eV), and effective mass (0.45$m_0$, $m_0$ is the free electron mass) are calculated for 1L $MoS_2$ using DFT. To keep the analysis simple, these quantities are assumed to be the same for different semiconductor thickness. Even after taking all possible precautions mentioned here, TCAD simulations have their limitations especially because they do not capture the interactions between metal and semiconductor at the interface (for additional discussion please refer to Appendix A). Nevertheless, these simulations can provide an accurate relationship between device electrostatics, current crowding, and $R_C$, which is one of the main subjects of this paper.

**Transmission Line Method (TLM):** To have an unambiguous definition of the contact resistance, virtual transmission line method (TLM) experiments are performed by simulating structures of various channel lengths (L) from 100 nm to 1 μm. The smallest channel length is kept larger than the channel thickness (L > $t_{2D}$) to avoid fringing effects. In these simulations, a small voltage ($V_{DD}$ = 0.1 V) is applied across the channel and the current through the device is calculated ($I_D$). To calculate the contact resistance, total device resistance ($R_T$ = $V_{DD}/I_D$) is plotted against respective L. The contact resistance is then calculated by fitting a line to this plot of $R_T$ versus L. The intersection point of this fit line with the y-axis (ordinate) gives 2$R_C$, where $R_C$ is the contact resistance.

### C. Experimental details

For the experimental extraction, we fabricate TLM structures with 1L $MoS_2$ grown by chemical vapor deposition (CVD) [28], [41]. First, the contacts are made using lithography and 40 nm thick Au deposited with electron-beam physical vapor deposition in high vacuum (~$10^{-8}$ Torr). The $MoS_2$ is then etched for well-defined widths using $O_2$ plasma and photoresist as an etch-mask. Each TLM structure has multiple devices with varying channel lengths from 100 nm to 2 μm. For one of the TLM structures, we cap the $MoS_2$ channel with sub-stoichiometric aluminum oxide ($AlO_x$) to induce n-type doping in the $MoS_2$ [42]. We use these TLM structures to extract the contact resistance for various back-gate voltages (and the carrier densities).

# Appendix

## A. Limitations of TCAD simulations

At the nanoscale, the contact metal interacts with the 2D materials [14]. It has been observed that the 2D material below the contact is significantly affected due to the interaction with the metal [20]. The metal could induce metallic states in the semiconductor and make the 2D material at the interface completely metallic and no band gap [19]. As a result, especially for few layers of 2D materials, the potential seen by the carrier at the interface is not accurate in Sentatarus TCAD. If such subtleties are not considered, it might lead to incorrect conclusions about carrier injection [13]. In experiments, there are additional factors such as surface impurities, metal contact grains which would further affect the carrier injection.

## B. Impact of semiconductor traps and band states on current injection

In Fig. 9a, we show an energy band structure below the contact after applying a certain back-gate voltage. Especially for very thin semiconductors, the conduction band may not bend sufficiently to have empty states for electrons from the metal to tunnel too as shown in Fig. 9a. As such the tunneling path for carriers shown by the red dotted arrow in Fig. 9a is forbidden. In other words, the carriers cannot be injected from the top contact, and the only possible path for carrier injection is through the contact edge. In realistic devices with semiconductor traps or metal induced gap states [20], the semiconductor below the contacts will have band gap states. We show in Fig. 9a (green arrow), that in the presence of such trap states we can still inject the carriers in the semiconductor from the top contact. Note that even though we did not consider traps for our previous simulations, our conclusions for Fig. 1 and 2 still hold because in all these cases $\rho_C$ and as a result $L_T$ are very small.

We show in Fig, 9b the contact resistance for three different cases as a function of channel thickness. The contact resistance for thermionic only case (brown diamond symbols) is highest, while the contact resistance when we allow tunneling is lowest ($\rho_C$ is low, blue square symbols). Adding trap (green square symbols) increases the contact resistance slightly due to a slight degradation in $R_{sh}$. In Fig. 9c-e, we show the current density below the contact for these three cases. As expected in Fig. 9c-d, the highest current density is observed close to the oxide-semiconductor interface and larger carrier injection seems to happen close to the edge. While in Fig. 9e, where we include traps in the semiconductor, we see that the larger carrier density is seen closer to the contact and the carrier injection is uniform throughout the contact length. In summary, in realistic semiconductors, it is more likely for the carriers to undergo top injection even for extremely thin channels.

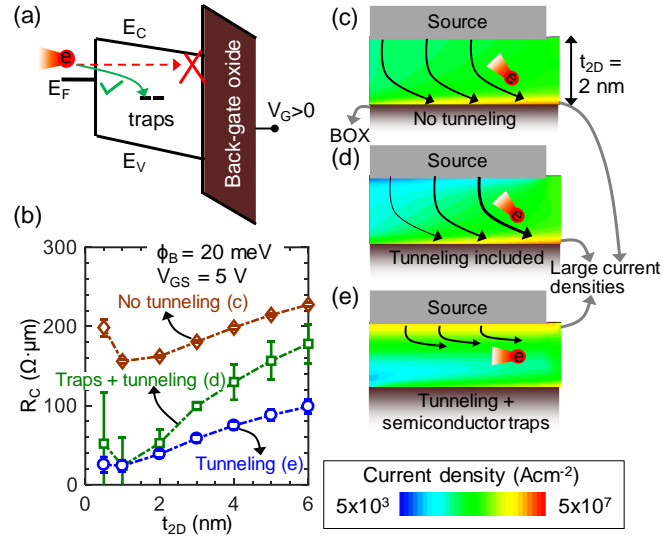

**Figure 9: Impact of traps on current injection.** (a) The energy band diagram in the semiconductor below the contact. (b) Comparison of the contact resistance with different carrier injections: only thermionic (brown diamond symbols), tunneling at the contact (blue circle symbols), and semiconductor traps in addition to tunneling (green square symbols). We have shown the current densities below the contact for (c) Schottky contact but all transport is thermionic as the non-local tunneling model (NLM) is switched off, (d) Schottky contact with NLM (tunneling) switched-on, and (e) Schottky contact with NLM (tunneling) and semiconductor traps. For these simulations, we have assumed $t_{BOX}$ = 5 nm and $V_{GS}$ = 5 V. We assume a n-type doping density of $N_D = 10^{19}$ cm$^{-3}$, $\phi_B$ = 20 meV, in-plane mobility of 50 cm$^2$V$^{-1}$s$^{-1}$, anisotropy of 10, $t_{2D}$ = 2 nm, and $L_C$ 50 nm.

## Acknowledgments

This work has been partly supported by the NCN-NEEDS program, which is funded by the NSF contract 1227020-EEC and by the Semiconductor Research Corporation (SRC). The study was also partly supported by the NSF EFRI 2-DARE grant 1542883, by the AFOSR grant FA9550-14-1-0251, and by the Systems on Nanoscale Information fabriCs (SONIC, one of six SRC STARnet Centers sponsored by MARCO and DARPA). K.K.H.S. acknowledges partial support from the Stanford Graduate Fellowship (SGF) program and NSF Graduate Research Fellowship under Grant No. DGE-114747.